\documentclass[10pt,letterpaper]{article}

\usepackage{cogsci}
\usepackage{graphicx}

\cogscifinalcopy 

\usepackage[
  style=apa,
  natbib=true,
  annotation=false,
]{biblatex}
\usepackage{float} 

\usepackage{xcolor}
\usepackage{soul}
\usepackage{hyperref}

\title{What makes an action sequence enjoyable to watch?}

\author[1]{\mbox{Jean-Peïc Chou*}}
\author[2]{\mbox{Kristine Zheng}} 
\author[2]{\mbox{Junyi Chu}} 
\author[1]{\mbox{Maneesh Agrawala}} 
\author[1,2]{\mbox{Judith E. Fan}}
\affil[1]{Department of Computer Science, Stanford University, United States}
\affil[2]{Department of Psychology, Stanford University, United States}
\affil[*]{\href{mailto:jeanpeic@stanford.edu}{\texttt{jeanpeic@stanford.edu}}}

\begin{document}

\maketitle

\begin{abstract}

People often seek out ways to watch others perform complex action sequences (e.g., sports).
What makes some sequences more enjoyable to watch than others?
We generated 24 video clips of gameplay from a Flappy Bird-style video game. Clips varied in difficulty (how often players succeeded on average) and in moment-to-moment uncertainty (how likely the player was to crash at any given step).
Participants ($N=864$) rated each video on one of three dimensions: how much they enjoyed it, how difficult the level appeared, or how dangerous the player's trajectory appeared.
We found that participants preferred videos where the player seemed to be completing more difficult obstacle courses, but dangerousness did not predict enjoyment ratings.
These findings show how procedurally generated stimuli can isolate the factors that affect how enjoyable an action sequence is to watch.


%

\end{abstract}

\section{Introduction}


On January 25, 2026, over six million people tuned in on Netflix to watch Alex Honnold ascend the 1,667-foot Taipei 101 tower—bare-handed, without ropes, and in real time. 
Spectators weren't learning to climb themselves, nor did the outcome affect their lives in any practical way. Rather, they watched for enjoyment. 
What makes some performances more enjoyable to watch than others?

Survey studies have found that spectators enjoy watching "close contests," "amazing feats," and "skillful plays" (\cite{pirouz_creating_2015, pizzo_esport_2018, zillmann_enjoyment_1989}).
These observations suggest a variety of features that may explain enjoyment, including how difficult a task is or how close to failure the performer seems moment to moment.
However, it remains unclear what information observers use to form these enjoyment judgments, whether such judgments are shared across viewers, and whether they track measurable properties of the events themselves. 

%

One framework linking affective responses to stimulus features comes from appraisal theories of emotion, which propose that feelings arise from how people interpret events (\cite{ellsworth2003appraisal}).
Such appraisals can range from simple evaluations, such as stimulus novelty or uncertainty, to more complex judgments about personal relevance or threat (\cite{Moors2013, smith2012affect}).
This framework predicts that how observers interpret an action should affect what they feel while watching it.

A long history of research in cognitive science has found that action understanding involves attending to both the static features of an environment and the moment-to-moment dynamics of actions (\cite{dennett_intentional_1987, heider_experimental_1944, wellman_understanding_2002}).
Adults, children, and infants are sensitive to the physical constraints that increase action costs (e.g., climbing steeper hills; \cite{gergely_teleological_2003, liu_ten-month-old_2017}) and danger (e.g., jumping over deeper trenches; \cite{gjata_what_2022, liu_dangerous_2022}), and they expect agents to minimize such costs. 
People also use differences in action trajectories to infer others' goals, beliefs, and competence (\cite{jara-ettinger_naive_2016, jara-ettinger_naive_2020, liu_how_2026}). For example, inefficient behaviors may signal ignorance (\cite{baker_action_2009}) or incompetence (\cite{leonard_who_2019}).
These evaluations extend beyond social inferences. Reasoning about external constraints and action dynamics also informs judgments about task difficulty (\cite{gweon_reverse-engineering_2017, yildirim_explaining_2019}) and 
interestingness (\cite{holdaway_measuring_2021, martinez_measuring_2023}).

\begin{figure*}[t]
  \centering
  \includegraphics[width=\textwidth]{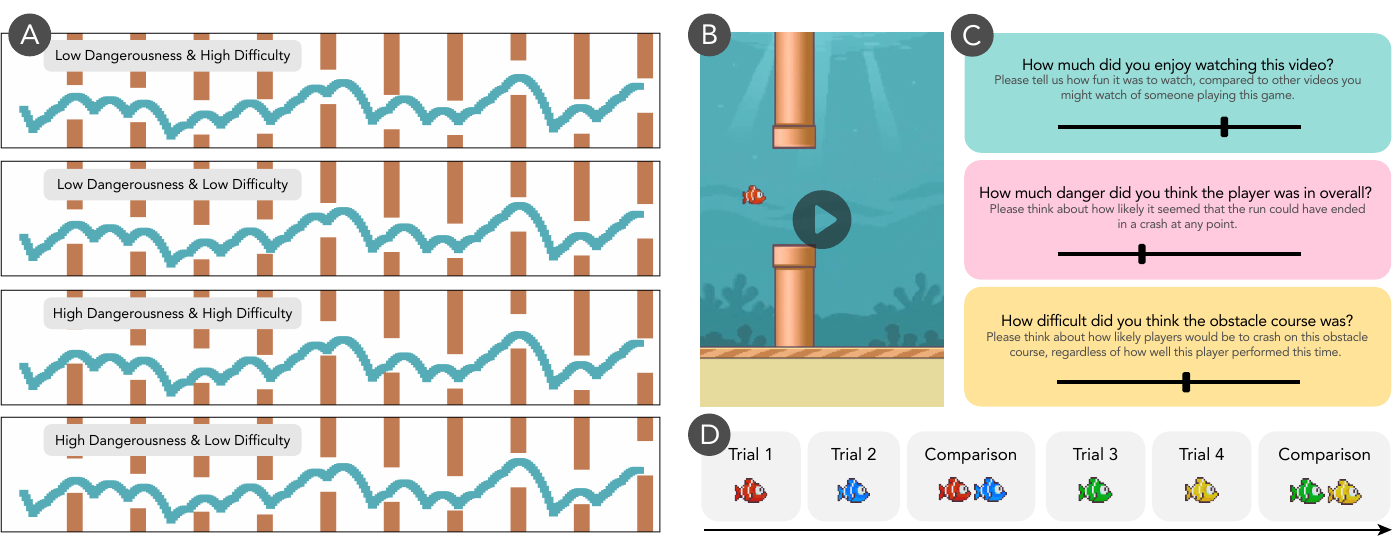}
  \caption{(A) Schematic visualizations of four stimuli showing the key experimental manipulation (Map difficulty $\times$ Dangerousness). Each example shows the same base trajectory (in blue, moving from left to right) in different obstacle layouts (in brown). Difficulty and dangerousness scores were computed using agent-based simulations (see Methods). Obstacle courses with higher difficulty scores have a higher rate of failure across simulated trajectories. Higher dangerousness scores reflect trajectories with smaller margins for error, meaning they would be more likely to result in collision under noisy perception or imprecise motor control. (B) A single video frame. (C) Participants rated each video on one of three dimensions using a 0--100 slider. The type of rating task was manipulated between groups of participants. (D) Study procedure. Participants viewed four gameplay videos (one per Dangerousness $\times$ Map difficulty condition) and rated each for either enjoyment, dangerousness, or difficulty.}
  \label{stimuli}
\end{figure*}

Here, we investigate whether the same dimensions that observers use to understand actions also predict their subjective feelings of enjoyment. As a first step, we consider two salient features: how challenging the environment is, and the moment-to-moment dangerousness of the agent's trajectory.

We developed and tested several competing hypotheses for how difficulty and dangerousness might affect how much viewers enjoy a performance.
First, viewers might enjoy performances situated in more difficult environments. When success seems improbable, observers may derive pleasure from witnessing novel, unlikely outcomes (\cite{pirouz_creating_2015}) or from appreciating the skill required to overcome challenges (\cite{pizzo_esport_2018, barney_exploration_2023}). 
Second, viewers may prefer performances with greater moment-to-moment uncertainty, as when agents execute risky maneuvers or navigate precarious situations. While many theoretical accounts treat the experience of uncertainty as inherently aversive (\cite{madrigal_effect_2011}), observing the resolution of such momentary dangers may be rewarding (\cite{kaspar_thrilling_2016, moulard_role_2019}).
Third, the two factors may interact: dangerous maneuvers may be especially enjoyable to watch in high-difficulty environments, where such danger is necessary for success, but reduce enjoyment in low-difficulty settings.
When safer paths are available, unnecessarily dangerous movements may suggest avoidable errors and imply lower skill, thus making the performance less enjoyable to watch.
Finally, uncertainty about agent trajectories may not specifically increase enjoyment but instead amplify emotional intensity, regardless of valence (\cite{bar-anan_feeling_2009}).
On this account, more dangerous trajectories should elicit more intense emotional responses --- positive or negative --- and produce greater spread or bimodality in self-reported enjoyment.
While these four hypotheses are not mutually exclusive, they make different predictions about the relative contributions of difficulty and dangerousness to enjoyment.

We test these hypotheses in a video game environment (Figure \ref{stimuli}) modeled after Flappy Bird, a simple single-player game that has attracted millions of online spectators (\cite{dredge_pewdiepie_2015, google2014_flappybird_yis}). We generated gameplay videos by simulating player agents, allowing us to independently manipulate environment difficulty and trajectory dangerousness while holding visual features and outcomes constant (Figure \ref{stimuli}A). Participants viewed these videos and rated each on apparent difficulty, apparent dangerousness, or their own enjoyment.

This design lets us directly test how difficulty and dangerousness influence subjective enjoyment. Our analysis focuses on three questions:
(1) the extent to which observers agree in their judgments of enjoyment, difficulty, and dangerousness;
(2) the degree to which subjective judgments of dangerousness and difficulty align with model-based estimates derived from simulated agent behavior; and most importantly, 
(3) the extent to which apparent dangerousness and apparent difficulty predict enjoyment, and whether these features interact. 

\begin{figure*}[t]
  \centering
  \includegraphics[width=\textwidth]{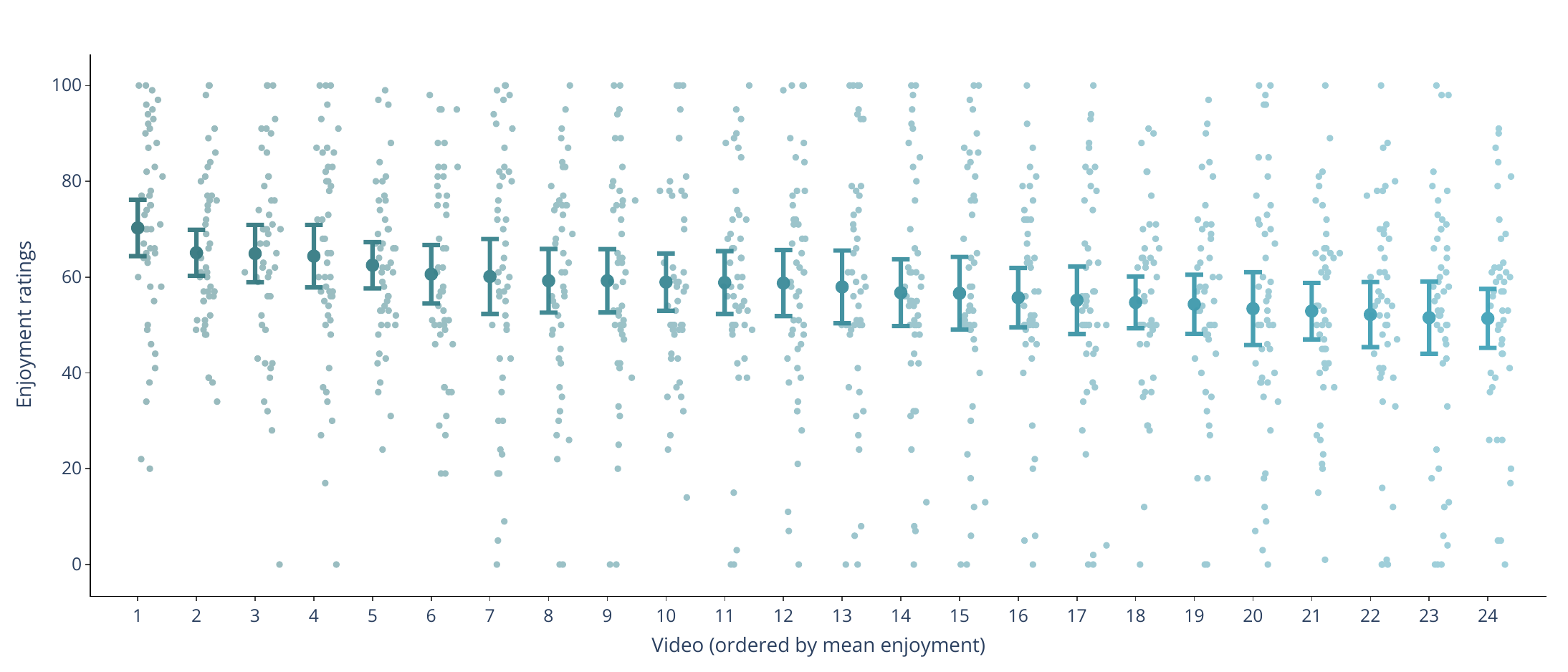}
  \caption{Distribution of enjoyment ratings for each of the 24 videos from highest to lowest mean enjoyment rating. Larger dots and error bars show the mean and 95\% CI; smaller dots show individual ratings (48 per video).}
  \label{enjoyment_ratings}
\end{figure*}

\section{Methods}
\subsection{Participants}

We recruited 905 US adults from Prolific (\textit{age}: median = 40; \textit{gender}: 44.2\% female, 53.7\% male). We excluded 41 participants for incomplete test trials, leaving a final analyzed sample of 864 participants, evenly assigned across the dangerousness, map difficulty, and enjoyment rating conditions ($N = 288$ per condition). The study took approximately five minutes to complete, and participants received \$1.50 USD in compensation.



\subsection{Stimuli}

We created 13-second videos from a custom variant of the game Flappy Bird, starting with a 3-second countdown followed by 10 seconds of gameplay footage. In this game, players navigate a fish through vertical obstacles ("pipes") by executing discrete upward "jumps", while facing constant downwards gravity and constant horizontal velocity. A run ends if the fish collides with an obstacle pipe, the ground, or the top of the screen.

To control for outcome, all videos show the fish successfully navigating exactly 10 consecutive obstacles. We generated six unique trajectories and created four videos with each trajectory, following a 2$\times$2 factorial design (low/high dangerousness $\times$ low/high difficulty). This design lets us decouple dangerousness from difficulty while controlling for trajectory shape (see Figure~\ref{stimuli}A).

\subsubsection{Operationalizing difficulty and dangerousness.} 
Our goal was to independently manipulate two stimulus features: how difficult the environment is overall, and the moment-to-moment dangerousness of an agent's trajectory.
Both features are defined relative to an agent's abilities. For example, a player with slower reaction time may be more likely to collide with nearby obstacles.
Because both features depend on agent ability, we quantified them by aggregating performance statistics across simulated agents with varying sensorimotor abilities.

We trained three reinforcement-learning agents to play the game under different levels of sensorimotor impairment (Low, Medium, High). Impairment combined perceptual noise (i.e., Gaussian noise on the perceived position of upcoming pipes) and motor noise (i.e., variable jump execution delay, sampled from a normal distribution), with both increasing across impairment levels. 
Agents were trained using Proximal Policy Optimization (PPO; \cite{schulman_proximal_2017}) to maximize survival time, yielding control policies adapted to their sensorimotor limitations.

We operationalized \textbf{dangerousness} as susceptibility to imminent failure. 
For each state in a trajectory, we queried the agent's learned value function, which estimates expected remaining survival time. Because our reward structure equates survival with cumulative reward, low values indicate states from which the agent does not expect to survive long. We averaged the negated value across the full trajectory so that higher scores indicate greater dangerousness.


We operationalized \textbf{difficulty} as the probability of failing to complete an obstacle course (\cite{isaksen_simulating_2017, gudmundsson_human-like_2018}). For each obstacle layout, we estimated failure probability from 1,000 simulated runs per agent, then averaged across all three agents.

\subsubsection{Generating videos.}
Using these dangerousness and difficulty scores, we created 24 videos following a factorial design, systematically varying difficulty and dangerousness while controlling for trajectory shape and outcome (Figure \ref{stimuli}A). All videos showed successful completion of exactly 10 obstacles.
 
We first identified six base trajectories with distinct movement patterns. From 1,000 simulated runs per agent, we selected the trajectories with lowest and highest mean vertical movement amplitude within each agent (3 agents $\times$ 2 amplitudes = 6 trajectories). This provided diversity in movement dynamics. 

For each base trajectory, we generated potential obstacle layouts that varied in difficulty and dangerousness. 
We sampled 1,000 collision-free layouts by randomly varying pipe heights around the trajectory. This established the feasible range of dangerousness and difficulty scores for each trajectory. 
We then selected four layouts per trajectory using extreme percentile thresholds ($\leq$ 5th, $\geq$ 95th), creating a 2$\times$2 factorial design (low/high dangerousness $\times$ low/high difficulty).

\begin{figure*}[t]
    \begin{center}
      \includegraphics[width=\textwidth]{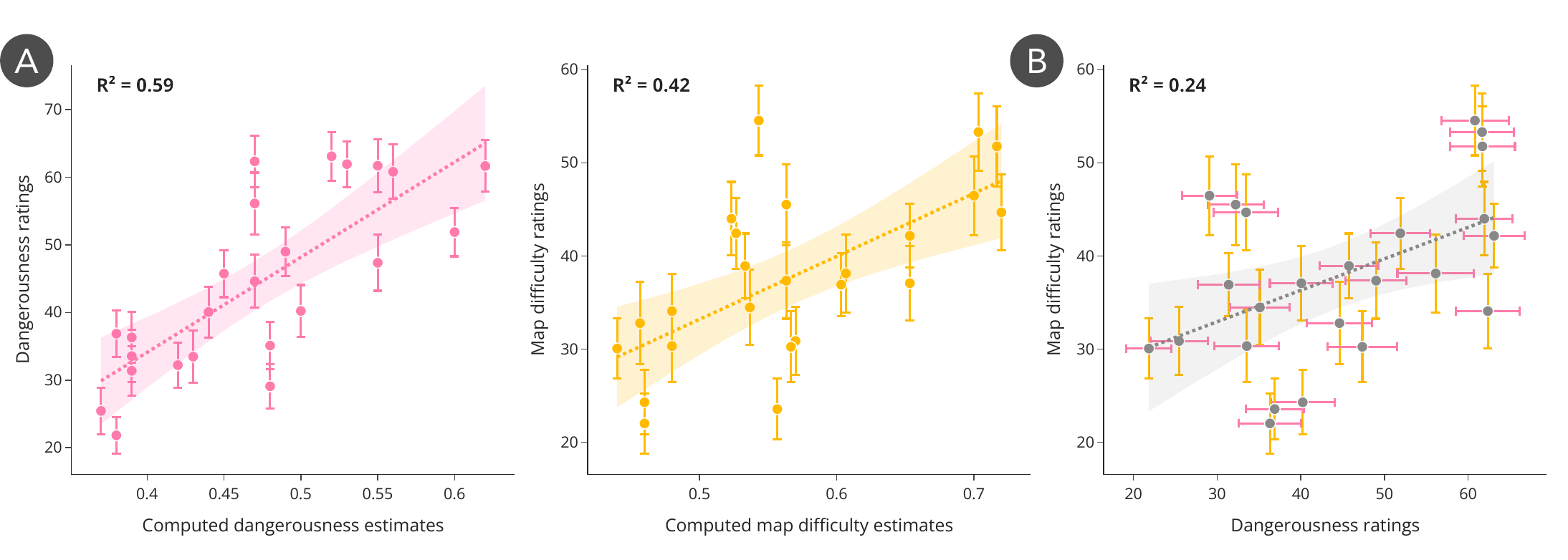}
  \end{center}
  \caption{(A) Correlations between ratings and model-based estimates for dangerousness and map difficulty. Points show per-video mean ratings. Error bars show 95\% bootstrapped CIs. Best fit lines show linear estimate with 95\% CI. (B) Correlation between map difficulty and dangerousness ratings.}
  \label{results1}
\end{figure*}

\subsection{Procedure}

Participants first read instructions explaining the game rules and watched a single example video --- gameplay in an extremely easy environment (i.e., large obstacle spacing) --- to familiarize themselves with the dynamics. They were then randomly assigned to one of three rating conditions (dangerousness, difficulty, or enjoyment), each with condition-specific instructions (Figure \ref{stimuli}C).

Each participant viewed four videos in sequence, one from each stimulus condition (high/low dangerousness $\times$ high/low difficulty).
Videos were presented in two blocks of two. The two videos within a block shared a base trajectory but were rendered on different obstacle courses, and together the four videos spanned all four (dangerousness $\times$ difficulty) conditions.
Each video featured a different colored fish (red, blue, green, and yellow) to help participants distinguish between them. The assignment of stimulus condition to video order and the mapping of fish colors to video order were both randomized across participants.


After watching each video, participants rated it on their assigned dimension using a continuous slider (0--100). Following each pair of videos, participants completed a forced-choice comparison, indicating which of the two videos scored higher on their assigned dimension. After completing all ratings, participants answered a brief post-study survey in which they explained their judgments.

\begin{figure*}[t]
    \begin{center}
      \includegraphics[width=\textwidth]{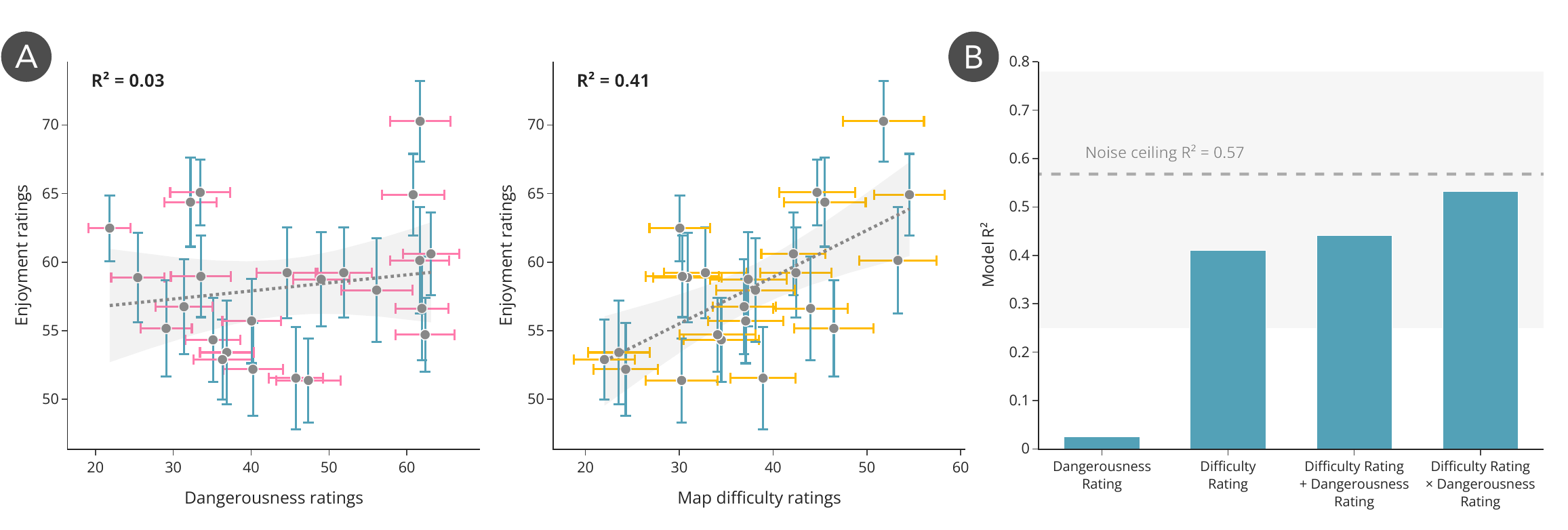}
  \end{center}
  \caption{(A) Correlations between enjoyment and dangerousness and map difficulty ratings. Each point represents the mean rating across participants for one video. Error bars show 95\% bootstrapped CIs. Best fit lines show linear estimate with 95\% CI. (B) Amount of variation in enjoyment that is attributable to ratings and trajectory. Y-axis show adjusted R² values from linear mixed effects models with different predictor terms, indicated on the X-axis. Noise ceiling shows maximal achievable R², estimated from the observed between-participant consistency in enjoyment ratings. Grey area indicates the noise ceiling 95\% CI.}
  \label{results2}
\end{figure*}

\section{Results}

\subsection{Viewers reliably judge enjoyment, difficulty, and dangerousness}

We first assessed whether videos elicited consistent judgments across participants.
We assessed stimulus-level reliability using a permutation-based split-half procedure with Spearman--Brown correction over 10{,}000 random splits (\cite{parsons_splithalf_2021}).

Participants showed strong agreement when judging apparent dangerousness ($r_{SB} = 0.93$, 95\% CI $[0.87,\,0.96]$) and map difficulty ($r_{SB} = 0.83$, 95\% CI $[0.71,\,0.91]$).
Participants showed moderate agreement in enjoyment ratings ($r_{SB} = 0.57$, 95\% CI $[0.25,\,0.78]$). Per-video enjoyment ratings appeared broadly unimodal, with disagreement expressed as spread rather than clear separation into groups (mean SD across videos = 22.3). Nevertheless, enjoyment ratings varied across stimuli with mean ratings ranging from $51.4$ to $70.3$ (see Figure~\ref{enjoyment_ratings}).

\subsection{Difficulty and dangerousness judgments track model-based estimates, but difficulty judgments are also influenced by dangerousness}

Next, we assessed whether self-reported judgments aligned with computed estimates derived from agent-based models. 

Dangerousness ratings closely tracked dangerousness estimates ($r(22) = 0.79$, $p < 0.001$), indicating that on average, the judgments aligned with estimated susceptibility to failure (see Figure~\ref{results1}A). Forced-choice comparisons within participant provided converging evidence.
Forced-choice judgments of which player was in more danger were near chance when both clips shared the same dangerousness condition (high-dangerousness \& high-difficulty vs. high-dangerousness \& low-difficulty: 53\% vs 47\%; low-dangerousness \& high-difficulty vs. low-dangerousness \& low-difficulty: 51\% vs 49\%; $N=96$ each). When a high-dangerousness clip was paired with a low-dangerousness clip ($N=384$), participants chose the high-dangerousness clip on 80\% of trials.

Mean map difficulty ratings also correlated strongly with computed difficulty estimates ($r(22) = 0.64$, $p < 0.001$; see Figure~\ref{results1}A).
However, difficulty judgments were also significantly associated with dangerousness, correlating with both mean dangerousness ratings ($r(22) = 0.46$, $p < 0.05$; see Figure~\ref{results1}B) and computed dangerousness estimates ($r(22) = 0.61$, $p < 0.01$).
Pairwise choices within participants corroborated these results: when high-dangerousness \& low-difficulty videos were paired with low-dangerousness \& high-difficulty videos ($N=96$), participants selected each clip as ``more difficult'' about equally often (51\% vs 49\%). In comparison, aggregating across all pairwise matchups, high-dangerousness \& high-difficulty and low-dangerousness \& low-difficulty videos were consistently judged most and least difficult, respectively. This pattern suggests that both obstacle layout and trajectory-level dangerousness shaped difficulty judgments.

\subsection{Viewer enjoyment ratings track judgments of map difficulty but not dangerousness}

Having established that judgments of dangerousness and map difficulty are reliable across viewers and track their computed estimates, we next tested whether these judgments predict enjoyment.

Videos reported as more difficult were also rated as more enjoyable ($r(22) = 0.68$, $p < 0.001$, $95\%~\text{CI} = [0.32, 0.83]$). By contrast, dangerousness ratings showed no relationship with enjoyment ($r(22) = 0.11$, $p = 0.62$; see Figure~\ref{results2}A).
A multiple regression using both factors to predict average enjoyment across the 24 videos\footnote{We also tested mixed-effects models, but with only four ratings per participant, the data structure did not support stable estimation of participant-level random effects.} explained almost half the observed variance ($R^2 = 0.44$). Difficulty rating was a strong positive predictor of enjoyment, whereas adding dangerousness rating as a main effect did not significantly improve fit over difficulty alone (nested model comparison: $F(1,21)=1.18$, $p = 0.29$; see Figure~\ref{results2}B). Adding a difficulty $\times$ dangerousness interaction term increased model fit ($R^2 = 0.53$) without reaching statistical significance (nested ANOVA: $F(1, 20)=3.84$, $p=0.06$). We observed a positive interaction ($\beta = 0.34$, $p = 0.06$), suggesting that the enjoyment boost from higher difficulty was larger when dangerousness was also higher. We treat this interaction as exploratory given the marginal $p$-value and the small stimulus set ($N=24$).

Next, we examined whether apparent dangerousness influenced the spread of enjoyment responses. 
We found no evidence for such an effect: dangerousness ratings were only weakly correlated with the per-video standard deviation of enjoyment ratings, and this correlation was not statistically significant ($r(22)=0.23$, $p=0.29$).

On average, viewers derived more enjoyment from performances on tasks they judged to be difficult, while apparent dangerousness had no consistent effect on enjoyment.

\section{Discussion}

People often enjoy watching others in action, but the event features that drive this enjoyment are not well understood. Appraisal theories of emotion suggest that subjective feelings depend on how individuals evaluate the events they experience, while a long history of work on action understanding has shown that people readily extract information about both static environmental constraints and the moment-to-moment dynamics of agents' actions.
Integrating these perspectives, we identified two candidate event features: the overall likelihood that an agent will succeed in a given environment, and the agent's susceptibility to failure at any given moment along a trajectory.
We tested how people respond to these features using simulated gameplay videos that independently manipulated environment difficulty and trajectory-level dangerousness.

We found that video enjoyment ratings were predicted by environment difficulty, with no consistent effect of trajectory-level dangerousness.
This pattern suggests that viewers primarily derive enjoyment from performances that overcome difficult environmental constraints, rather than from the moment-to-moment dangers agents encounter.
Notably, this pattern held even though dangerousness influenced difficulty judgments themselves, suggesting that viewers extract and weight environmental difficulty independently of dangerousness when judging enjoyment.

The null effect of dangerousness is striking given that nearly half of participants spontaneously mentioned dangerousness-related factors in a post-study survey explaining their responses.
One possibility is that dangerousness mattered to some degree but was masked by measurement noise due to individual differences.
For example, people differ in their general propensity to feel positive emotions (\cite{hamilton1984intrinsic}) and in their tolerance for risk and uncertainty (\cite{weber2002domain}). Participants may also interpret ``enjoyment'' in diverse ways (e.g., as excitement, interest, or suspense), producing heterogeneous responses to the same stimuli.
Consistent with this account, inter-rater reliability was lower for enjoyment than for difficulty or dangerousness judgments.
Future work should collect more ratings per participant across a broader set of evaluative dimensions, which would help separate individual differences in trait-level enjoyment from inconsistency in how ``enjoyment'' is interpreted.

However, the contrast between participants' explicit explanations and our experimental results may reflect more than just measurement noise. Instead, the contrast may reflect a substantive dissociation: people's intuitions about what drives their enjoyment may diverge from the features that actually do.
This dissociation points to the limits of introspection as a guide to aesthetic preference, and to the need for measures of enjoyment beyond self-report.
In particular, future research should complement self-reports with more implicit measures of enjoyment, including physiological and behavioral measures of emotion and attention (e.g., pupillometry), and consider continuous measures to distinguish retrospective evaluations from enjoyment as it unfolds.




In this work, we chose the Flappy Bird environment because the constrained action space provided a controlled setting for dissociating environment-based difficulty and moment-to-moment trajectory dangerousness.
However, the correlations between perceived difficulty and dangerousness suggest that these factors are not independent appraisals. Viewers may use inferences about an agent's competence and intentionality when judging both the difficulty of the obstacle course and the dangerousness of the agent's actions. For example, a dangerous-looking trajectory may make an otherwise easy layout appear difficult if a competent player genuinely seems to struggle.
Distinguishing these interpretations requires environments in which players can pursue different trajectories through the same situation, which would allow observers to infer whether dangerousness reflects task demands, intentional risk-taking, or lack of competence. Our Flappy Bird environment is limited in this respect because the obstacle layouts strongly constrain the range of viable trajectories. Increasing the range of possible trajectories and playing styles, for example by widening obstacle gaps or leveraging games with richer action spaces, would make it easier to distinguish these interpretations.



Another limitation of the environment concerns potential confounds between task difficulty and trajectory shape.
Because difficulty is defined with respect to obstacle positions, larger height changes between successive pipes both increased difficulty and required agents to make larger vertical movements. 
Consistent with this possible confound, we found that on average, higher-amplitude trajectories were rated as both more difficult and more enjoyable.
This pattern raises the possibility that the observed difficulty-enjoyment association may partly reflect responses to movement dynamics rather than environmental constraints per se.
Future work should disentangle these factors by exploring a wider range of trajectory shapes.

Despite this potential confound, viewer judgments were strikingly consistent given the brief exposure: participants viewed just 10 seconds of movement through individual obstacles, yet consistently extracted both difficulty and dangerousness.
This rapid extraction likely depends on efficient perceptual processes: observers may track specific diagnostic cues such as obstacle spacing for difficulty or proximity to collision for dangerousness.
The selective influence of difficulty on enjoyment also suggests that evaluation involves more than perception alone: viewers may weight perceived features by their relevance to evaluative goals, a form of goal-directed attention (\cite{Borji2014}).
Future work using eye-tracking methods could reveal whether participants spontaneously attend to different features when judging enjoyment versus difficulty.
More broadly, our approach illustrates the value of using procedurally generated stimuli to study action evaluation.
By systematically manipulating environment difficulty and trajectory dangerousness within different movement patterns, we could test precise mechanistic hypotheses about what makes a performance enjoyable to watch.
Our operationalizations --- difficulty as the expected likelihood of success on a task, and dangerousness as the susceptibility to failure along a specific trajectory --- could be extended to domains beyond simple arcade games, from board games to multi-agent settings (e.g., \cite{collins_people_2025}).



Together, these results identify specific event-level features --- environmental difficulty in particular --- that predict spectators' affective responses, and link the perception of action constraints to the affect they produce. Enjoyment of unfolding action is not ineffable: it tracks observable features of the events being watched.

\section{Acknowledgements}
We thank Ian-Christopher Tanoh and members of the Cognitive Tools Lab at Stanford University for helpful discussion.

This work was supported by awards from the Brown Institute for Media and Innovation to J.P.C. and J.E.F.; NSF CAREER Award \#2436199, NSF DRL \#2400471 to J.E.F.; and awards from the Stanford Human-Centered AI Institute (HAI) and Stanford Accelerator for Learning to J.C. and J.E.F.

\section{Code availability}
Code and materials are available at \url{https://github.com/cogtoolslab/enjoyable-action-sequences-cogsci26}.

\printbibliography

\end{document}